\documentclass[aps,pra,twocolumn,showpacs]{revtex4-1}

\usepackage{graphicx}
\usepackage{color} 
\usepackage{bm}
\usepackage{braket}
\usepackage{amsmath, amsthm, amssymb,mathrsfs}
\usepackage{csquotes}
\usepackage[colorlinks=true, citecolor=blue,allcolors=blue]{hyperref}

\DeclareMathOperator{\acot}{acot}

\DeclareMathOperator{\atan}{atan}
\DeclareMathOperator{\sgn}{sgn}

\newcommand{\IR}{\scriptscriptstyle\textsc{IR}}
\newcommand{\XUV}{\scriptscriptstyle\textsc{XUV}}

\begin{document}

\title{Resonant Anisotropic Emission in Two-Photon Interferometric Spectroscopy}

\author{B.~Ghomashi$^1$, N.~Douguet$^{1*}$, and L.~Argenti$^{1,2*}$}
\affiliation{$^1$Department of Physics, University of Central Florida, Orlando, Florida 32186, USA}
\affiliation{$^2$CREOL, University of Central Florida, Orlando, Florida 32186, USA}
\date{\today}
\email{nicolas.douguet@ucf.edu}
\email{luca.argenti@ucf.edu}


\begin{abstract}
We theoretically explore a variant of RABBITT spectroscopy in which the attosecond-pulse train comprises isolated pairs of consecutive harmonics of the fundamental infrared probe frequency. In this scheme, one-photon and two-photon amplitudes interfere resulting in an asymmetric photoelectron emission. This interferometric principle has the potential of giving access to the time-resolved ionization of systems that exhibit autoionizing states, since it imprints the group delay of both one-photon and two-photon resonant transitions in the energy-resolved photoelectron anisotropy as a function of the pump-probe time delay. To bring to the fore the connection between the pump-probe ionization process and its perturbative analysis, on the the one side, and the underlying field-free scattering observables as well as the radiative couplings in the target system, on the other side, we test this scheme with an exactly solvable analytical one-dimensional model that supports both bound states and shape-resonances. The asymmetric photoelectron emission near a resonance is computed using perturbation theory as well as solving the time-dependent Sch\"odinger equation; the results are in excellent agreement with the field-free resonant scattering properties of the model.
\end{abstract}

\maketitle

 \section{Introduction} \label{sec:introduction}

Since its discovery more than a century ago~\cite{Hertz1887}, the photoelectric effect, i.e., the emission of an electron from an atom, molecule or extended target, due to the absorption of ionizing radiation, has played a fundamental role in our understanding of charge-transfer processes in matter. Photoionization often involves the excitation of transiently bound electronic states, which decay by emitting an electron on a time scale as small as a few femtoseconds~\cite{Wickenhauser2005,Gruson2016}. Until the end of the XX century, photoelectron spectroscopies were mostly limited to study photoemission processes in the stationary regime~\cite{Schmidt1992}. During the last two decades, however, the development of novel sources of sub-femtosecond extreme ultraviolet (XUV) light pulses has opened the way to the time-resolved study of electronic dynamics in atoms and molecules at its natural timescale~\cite{Hentschel2001,Paul2001,Sansone2006,Krausz2009}. Attosecond spectroscopy has allowed the measurements and theoretical computations of the photoionization time delay from different electronic valence shells~\cite{Schultze2010a,Kheifets10,Isinger17}, 
across autoionizing resonances~\cite{Argenti2010,Argenti2017,Galan16,Gruson2016,Cirelli18,douguet18}, or as a function of the emission angle~\cite{Comby16,Beaulieu16,Beaulieu17,Serov16,Cirelli18}. 

Reconstruction of Attosecond Beating by Interference of Two-Photon Transitions (RABBITT) \cite{Paul2001, Muller2002, Agostini2004} is a popular technique in which an XUV attosecond-pulse train (APT) is used as a pump, in conjunction with a weak infrared (IR) probe pulse, to ionize a target, as a function of the pump-probe delay $\tau$. The APT is obtained through the process of High-Harmonic Generation (HHG) \cite{LHuillier1993,Corkum1993}, which takes place when a strong replica of the IR probe pulse is focused on an active medium. 

The XUV pulse trains obtained with the HHG processes described above are generally weak (less than $10^{10}$~W/cm$^{2}$ on focus), and hence they can only be used to promote the absorption of a single harmonic XUV-photon by a target. When used in isolation, a pump APT ionizes the target giving rise to a photoelectron spectrum featuring sharp peaks in correspondence with the harmonics of the fundamental IR frequency. If the ionization takes place in the presence of a weak probe replica of the initial IR pulse, with a controllable delay with respect to the XUV train, however, two-photon ionization paths become possible \cite{Paul2001}. Besides simple one-photon absorption, the other most relevant ionization paths entail  also the exchange (emission or absorption) of one IR-photon. 

\begin{figure}[t]
\includegraphics[width=\columnwidth]{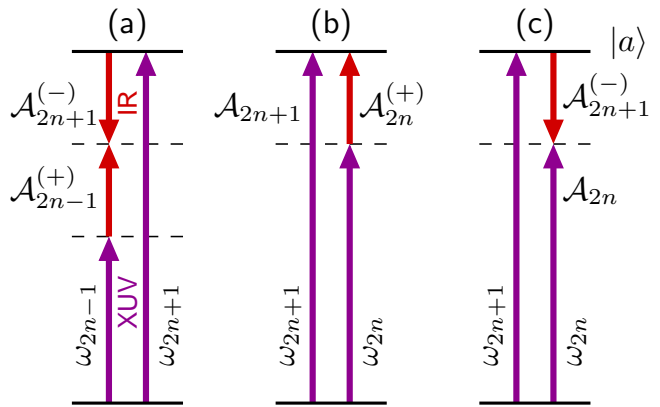}
\caption{\label{fig:1} RABBITT schemes used to study the dynamics of an autoionizing resonant state $|a\rangle$. (a) traditional RABBITT. (b) One-Two RABBITT for ionization paths interfering at the resonance, or (c) at the lower harmonic with the resonance used as an intermediate stepping stone (see text for details).}
\end{figure}

In traditional RABBITT schemes, the APT is linearly polarized and its spectrum comprises only odd multiples of the fundamental IR frequency, $\omega_{\IR}$. The absorption of an XUV photon from the $2n-1$ harmonics, followed by the absorption of one IR photon, and the absorption of an XUV photon from the $2n+1$ harmonics, followed by the stimulated emission of one IR photon, result in transition amplitudes to the same final energy in the continuum, $\mathcal{A}^{(+)}_{2n-1}$ and $\mathcal{A}^{(-)}_{2n+1}$, respectively, which give rise to the $2n$-sideband signal in the photoelectron spectrum, as  schematically illustrated in Fig.~\ref{fig:1}a. Thanks to the inherent coherence between the XUV and IR pulses, these two amplitudes interfere. Each exchange of an IR photon imparts to the ionization amplitude a phase factor $\exp(\mp i\omega_{\IR} \tau)$, depending on whether the photon is absorbed or emitted. As a consequence, the photoionization probability to the sideband $I_{2n}$ oscillates as a function of the pump-probe time delay $\tau$ at twice the frequency of the IR,  $I_{2n}\propto|\mathcal{A}^{(+)}_{2n-1}\,\mathcal{A}^{(-)}_{2n+1}|\cos(2\omega_{\IR}\tau+\phi_{2n})$~\cite{Veniard1996,Jimenez2014}, where $\phi_{2n}$ is a characteristic phase shift.
The phase shift $\phi_{2n}$ incorporates both the relative phase of two consecutive harmonics  and the additional phase imparted to the photoelectron by the two-photon transition itself. In the special case in which one of the two harmonics excites a metastable state $|a\rangle$, therefore, the other harmonic can serve as a holographic reference to measure the phase of the resonant two-photon transition~\cite{Argenti2010,Argenti2017,Galan16,Gruson2016,douguet18}. Indeed, the phase shift in the sideband beating exhibits the rapid excursion characteristic of resonant amplitudes as a function of the detuning of the first harmonic from the intermediate resonance. From the phase profile it is possible to reconstruct the fast decay of the metastable state to the continuum, resolved in time~\cite{Gruson2016}. 

The ability of this pioneering approach to study the time evolution of autoionizing states, however, is limited. In particular, the reconstruction of one-photon resonance decay is highly distorted because the resonance excitation amplitude is buried in a two-photon transition where it is inextricably intertwined with the contribution of virtual excitations to states across a wide range of energies. Furthermore, one-photon amplitudes are imaged to sidebands through structured continuum-continuum couplings that distort the amplitude phase as a curved mirror distorts the image of a visitor of an amusement park~\cite{Argenti2017}. Finally, the duration of IR probe pulses in attosecond experiments is typically of the order of tens of femtoseconds, i.e., comparable to or even shorter than the lifetime of most autoionizing states. As a result, the energy resolution of the already distorted resonant profile is limited by the energy resolution of the probe photon.

In the RABBITT setup illustrated above, when applied to spherically symmetric targets such as closed-shell atoms, the portions of the state function responsible for the harmonic and the sideband photoelectron amplitudes have well defined parity, i.e., odd and even, respectively, corresponding to the parity of the number of exchanged photons. As a consequence, the photoelectron distributions are symmetric upon reflection on the plane perpendicular to the polarization axis of the external pulses.
In the present work we propose an alternative interferometric scheme, which we call \emph{1-2 RABBITT},  that overcomes and quantifies the distortions inherent to the reconstruction of resonance decay with traditional RABBITT. In 1-2 RABBITT, the XUV spectrum comprises isolated pairs of linearly polarized consecutive harmonics of the probe frequency, as illustrated in Fig.~\ref{fig:1}.b. This scheme can be realized, for example, by employing an attosecond pulse train with two largely dominant consecutive odd harmonics, together with the second harmonic of the fundamental IR as a probe. Conversely, it is well known that if a small second-harmonic component is added to the strong IR pulse used in the HHG~\cite{Laurent2012, Mauritsson2008}, the resulting pulse train can comprise both even and odd multiples of the fundamental IR frequency. In the latter case, two consecutive harmonics could conceivably be isolated with the help of a suitable metallic filter. Regardless of the specific generation mechanism, for the sake of being definite, as well as to maintain a close parallel with traditional RABBITT schemes, in the following we will label the two consecutive harmonics as $2n$ and $2n+1$.
The one-photon ionization amplitude $\mathcal{A}_{2n+1}$, tuned to an isolated resonance $|a\rangle$, interferes with the two-photon amplitude $\mathcal{A}^{(+)}_{2n}$, giving rise to an asymmetric photoelectron emission along the laser polarization $\hat{\epsilon}$, $\Sigma(E)=\int d\Omega\, I_{2n}(E,\hat{\Omega})\sgn(\hat{\Omega}\cdot\hat{\epsilon})$, that beats at the IR frequency as $\Sigma(E)\propto|\mathcal{A}_{2n+1}\,\mathcal{A}^{(+)}_{2n}|\cos(\omega_{\IR}\tau+\phi_{2n+1})$, where $\phi_{2n+1}=\arg\mathcal{A}_{2n+1} - \arg\mathcal{A}^{(+)}_{2n}$. In contrast to the traditional RABBITT scheme, therefore, in this case the phase of the one-photon transition to the resonance $|a\rangle$ is directly encoded in $\phi_{2n+1}$, and can be extracted from it, provided that the two-photon amplitude is not structured. Furthermore, the same experiment quantifies the distortion of the resonant profile introduced by the probe photon. Indeed, at the harmonics $2n$, the amplitudes $\mathcal{A}_{2n}$ and $\mathcal{A}^{(-)}_{2n+1}$ interfere (see panel (c) of Fig.~\ref{fig:1}),
leading to a $\omega_{\IR}$-beating of $\Sigma(E)$ with terms of the form $\propto|\mathcal{A}_{2n}\,\mathcal{A}^{(-)}_{2n+1}|\cos(\omega_{\IR}\tau+\phi_{2n})$, where $\phi_{2n}$ now exhibits the excursion of the intermediate resonant phase modulated by a free-free transition induced by the IR field. 

In either case, the beating distribution has odd parity, since it originates from the interference between odd and even partial waves, and hence it can only be observed with a directional detector. On the other hand, the beating amplitude is proportional to the XUV intensity and to the peak IR field, which makes this scheme more sensitive than traditional RABBITT, where the signal is proportional to the IR intensity instead. Notice that the relative phase of the two consecutive harmonics is not relevant here, since the observable of interest is the rapid variation of the photoionization phase across the energy of a single harmonic, rather than its absolute value.

To illustrate the 1-2 RABBITT scheme we will employ a one-dimensional model that exhibits the key features of asymmetric resonant photoemission. The choice of a model whose attributes are analytically known allows us to better highlight the correspondence between scattering phase, photoionization phase, and the phase reconstructed from one-photon and two-photon resonant transitions, than with a many-body systems, in which the decay is driven by correlation. Furthermore, in the context of perturbative single-ionization processes, a 1D model is justified as it already reproduces most of the features of atomic ionization. This circumstance is to be contrasted with the case of strong-field ionization, where 1D models have a more limited validity since they cannot reproduce the transverse spreading of the electronic wavepacket, which is an essential aspect of all processes that depend on the recollision of the photoelectron with the parent ion.

The paper is organized as follows. In \hbox{Sec. \ref{sec:time-evolution}} we summarize the time-dependent formalism appropriate for 1-2 RABBITT, illustrate the one-dimensional model, and describe its relevant observables.
 In \hbox{Sec. \ref{sec:Results}}, we present the results for selected one-photon and two-photon transition amplitudes obtained, for several pump-probe time delays, using time-dependent perturbation theory (PT) as well as by integrating the time-dependent Schr\"{o}dinger equation (TDSE) numerically. The excellent agreement between these two approaches confirms that the analytical perturbative formulas capture all the aspects of the 1-2 RABBITT interferometric scheme. From the photoionization asymmetry computed with either methods, we retrieve the one-photon and the two-photon complex resonant ionization amplitude. In section {\ref{sec:CONCLUSION} we present our conclusions and perspectives.

\section{Theoretical approach} \label{sec:time-evolution}
\subsection{Time evolution}
Atomic units ($\hbar=1$, $e=1$, $m_e=1$) are used throughout, unless stated otherwise. The time evolution of the wavefunction $\Psi(t)$ of a quantum system driven by external time-dependent fields is governed by the TDSE,
\begin{eqnarray}
	i\frac{\partial \Psi(t)}{\partial t}=H(t)\Psi(t) \label{eq:TDSE},\\
	H(t)=H_0+H'(t) \label{eq:ham}
\end{eqnarray}
where the total Hamiltonian $H(t)$ has a field-free component, $H_0$, and a radiation-matter interaction component, $H'(t)$. In the 1D model examined in this work, $H_0=p^2/2+V(x)$, where $p=-i d/dx$ is the electron momentum and $V(x)$ is a local potential. In dipole approximation, the minimal-coupling interaction term is $H'(t)=F(t)o$, where $F(t)$ is the time-dependent field, and $o$ a suitable dipole operator. In velocity gauge $H'(t)=A(t)p$, where $A(t)$ is the vector potential, whereas in length gauge $H'(t)=-\mathcal{E}(t)x$, where $\mathcal{E}=-\partial_t A$ is the external electric field~\cite{Joachain}.
In the case of weak fields with finite duration, the solution of Eq. \ref{eq:TDSE} can be found as a truncated perturbative series. To express the perturbative solution, it is convenient to reformulate the TDSE in the interaction representation and in integral form,
\begin{equation}
	\Psi_I(t)=e^{iH_0t}\Psi(t),\qquad H^\prime_I(t) =e^{iH_0t} H^\prime(t) e^{-iH_0t},
\end{equation}\vspace{-15pt}
\begin{equation}
	\Psi_I(t)=\Psi_I(t_0)-i\int_{t_0}^t dt'H_I'(t')\Psi_I(t').
\end{equation}
The wave function is then the sum of finite-order terms, 
\begin{eqnarray}
	\Psi_I(t)&=& \sum_m\Psi^{(m)}_I(t),\\
	\Psi_I^{(m)}(t)&=& -i\int_{t_0}^t dt'H_I'(t')\Psi^{(m-1)}_I(t').
\end{eqnarray}
Let us assume that the field-free Hamiltonian has a non-degenerate ground state, $H_0|g\rangle=E_g|g\rangle$, normalized to unity, $\langle g | g \rangle = 1$, and a set of ionization channels $H_0|\alpha E\rangle=E|\alpha E\rangle$, normalized as $\langle \alpha E | \beta E'\rangle=\delta_{\alpha\beta}\delta(E-E')$, where $E$ is the total energy and $\alpha$ is a set of additional quantum numbers needed to resolve possible degeneracies. If the system is initially in the ground state,  the ionization probability amplitude $\mathcal{A}_{\alpha E\leftarrow g}$ to the final state $\ket{\psi_{\alpha E}}$ at any time $t$ after the end of the pulse is
\begin{equation}
\mathcal{A}_{\alpha E\leftarrow g }=\sum_{m}\mathcal{A}^{(m)}_{\alpha E\leftarrow g}, \quad \mathcal{A}^{(m)}_{\alpha E\leftarrow g}=\braket{\psi_{\alpha E}|\Psi_I^{(m)}(t)}.
\end{equation}
The first- and second-order ionization amplitudes can be expressed in the frequency domain~\cite{Jimenez2014} as
\begin{eqnarray}
\mathcal{A}_{\alpha E\leftarrow g}^{(1)}&=&-i\braket{\psi_{\alpha E}|o|\psi_g}\tilde{F}(\omega_{Eg}), \label{eq:one-photon}\\
\mathcal{A}_{\alpha E\leftarrow g}^{(2)}&=&-i\int_{-\infty}^{\infty}d\omega\tilde{F}(\omega_{Eg}-\omega)\tilde{F}(\omega)\mathcal{M}_{\alpha E\leftarrow g}^{(2)}(\omega)\label{eq:two-photon},
\end{eqnarray}
where the two-photon matrix element has the form
\begin{eqnarray}
\mathcal{M}_{\alpha E\leftarrow g}^{(2)}(\omega)=\braket{\psi_{\alpha E}|\,o\, G_0^{+} (E_g+\omega) \,o\,|\psi_g}\label{eq:two-photon-matrix},
\end{eqnarray}
$\tilde{F}(\omega)=1/\sqrt{2\pi}\int dt F(t)\exp(-i\omega t)$ is the Fourier transform of the field, $G_0^+(E)=(E-H_0+i0^+)^{-1}$ is the retarded resolvent of the field-free Hamiltonian, and $\omega_{Eg}=E-E_g$ is the overall excitation energy~\cite{Joachain}. If the contribution of intermediate bound states is negligible (as it is certainly the case if the potential supports only one bound state), to the two-photon transition is
\begin{eqnarray}
\mathcal{M}_{\alpha E\leftarrow g}^{(2)}(\omega)=\sum_\beta\hspace{-2pt}\int\hspace{-2pt}\frac{\langle \psi_{\alpha E}|o|\psi_{\beta E'}\rangle\langle \psi_{\beta E'}|o|\psi_{g}\rangle}{E_g+\omega-E'+i0^+}dE'.\quad\label{eq:two-photon-matrix}
\end{eqnarray}
From the bound-free $\langle \psi_{\alpha E}|o|\psi_g\rangle$ and free-free $\langle \psi_{\alpha E}|o|\psi_{\beta E'}\rangle$ dipole matrix elements, therefore, it is possible to determine the photoionization probability and asymmetry resulting from the interaction between the system and an arbitrary sequence of weak pulses, under the assumption that all terms beyond second order are negligible. An alternative way to compute the ionization amplitude, which is valuable as a proof of principle to validate the perturbative results, is to solve the TDSE numerically, as discussed in the next sections.

\subsection{Analytical 1D model}
We consider a symmetric analytical model (AM) in which the potential $V(x)$ is given by two repulsive delta functions, located at $x=\pm a$, and an attractive delta potential located at the origin.
\begin{equation}
V(x)=V_+\delta(x+a)+V_+\delta(x-a)-V_-\delta(x), \label{eq:pot_analy}
\end{equation}
Where $V_+$ and $V_-$ are positive parameters that express the strength of the potential.
\begin{figure}[t]
	\includegraphics[width=0.7\columnwidth]{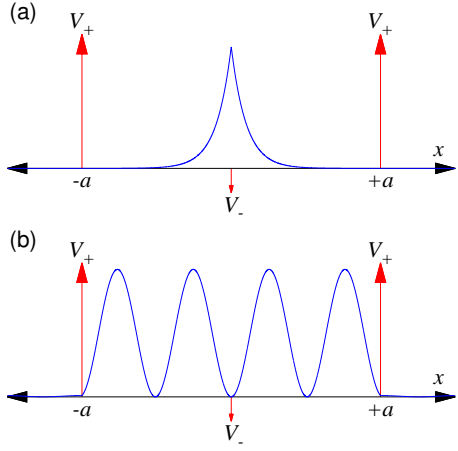}
	\caption{Analytical potential used in this study. Also shown the absolute value squared (a) of the ground state and (b) of a resonant state with odd parity.}
	\label{fig:2}
\end{figure}
\paragraph{Eigenstates}
The stationary states of the field-free Hamiltonian can be found analytically. Where the potential is zero the Hamiltonian eigenfunctions are 
\begin{eqnarray}
	\psi(x)&=&Ae^{\kappa x} + Be^{-\kappa x},\quad E<0,\quad \kappa =\sqrt{-2E},\\
	\psi(x)&=&Ae^{ikx} + Be^{-ikx},\quad E>0,\quad k =\sqrt{2E}.
\end{eqnarray}
The solution must be continuous, and its derivative satisfy the well known conditions 
\begin{eqnarray}\label{eq:boundary-conditions}
\psi'(0^+)&=&\psi'(0^-)-2V_-\psi(0),\\
\psi'(\pm a^+)&=&\psi'(\pm a^-)\,+\, 2\,V_+\psi(\pm a),
\end{eqnarray}
where $\pm a^{+}$ and $\pm a^{-}$ denote the right and left limit towards $x=\pm a$, respectively. Since the potential is symmetric about the origin, the parity operator $\Pi$, defined by $\Pi\ket{x}=\ket{-x}$, commutes with the field-free Hamiltonian $\left[H_0,\Pi\right]=0$. As a consequence, the eigenfunctions of $H_0$ can be assumed to be either even, $\psi_{eE}$, or odd, $\psi_{oE}$, under inversion of the spatial coordinate.

It is easy to ascertain that the potential~\eqref{eq:pot_analy} supports only one bound state.
If the repulsive potentials are far enough from the origin or, conversely, the attractive potential is sufficiently strong, 
the wavefunction of the bound state is negligible at $|x|\ge a$ and so is the discontinuity of its derivative.
For a whole range of parameters, therefore, the ground state is well approximated by that of a single attractive delta potential,
\begin{equation}
\psi_g(x)=A_ge^{{-\kappa |x|}},\quad E_g=-V_-^2/2,
\end{equation}
where $A_g=\sqrt{V_-}$ and $\kappa=V_-$. In particular, the ground state is an even function, as shown in Fig.~\ref{fig:2}a.
The even scattering functions are
\begin{equation}\label{eq:phi_even}
\hspace{-3pt}\psi_{eE}(x)=\frac{1}{\sqrt{\pi k}}\left\{
\begin{array}{ll}
A_e(k)\sin[k|x|+\delta_e(k)]&|x| < a\\\vspace{-8pt}\\
\sin[k|x|+\eta_e(k)]&|x|>a
\end{array}
\right.
\end{equation}
where $\delta_e(k)=-\atan{\left(k/V_-\right)}$ is the wave-function phase inside the barrier ($|x|<a$),
$\eta_{e}(k)$ is the scattering phaseshift, and $A_e(k)$ the asymptotic amplitude, 
\begin{eqnarray}
\eta_e(k)&=&\acot\left[\cot(ka+\delta_e)+2V_+/k\right]-ka\nonumber\\
        A_e(k)&=&\big[1+4V_+^2\,\sin^2[ka+\delta_e(k)]/k^2+\nonumber\\
        &&+2V_+\,\sin[2ka+2\delta_e(k)]/k\big]^{-1/2}.\nonumber
\end{eqnarray}
The odd scattering functions are
\begin{equation}\label{eq:phi_odd}
\hspace{-3pt}\psi_{o E}(x)=\frac{1}{\sqrt{\pi k}}\left\{
\begin{array}{ll}
A_o(k)\sin(kx)&|x| < a\\\vspace{-8pt}\\
\sin[kx+\sgn(x)\,\eta_o(k)]&|x|>a
\end{array}
\right.
\end{equation}
whose scattering phase and asymptotic amplitudes are
\begin{eqnarray}
\eta_o(k)&=&\acot\left[\cot(ka)+2V_+/k\right]-ka,\nonumber\\
        A_o(k)&=&\big[1+4V_+^2\sin^2(ka)/k^2+2V_+\sin(2ka)/k\big]^{-1/2}.\nonumber
\end{eqnarray}
Notice that the scattering states in~(\ref{eq:phi_even},\ref{eq:phi_odd}) are purely real, $\psi_{e/o\,E}^*(x)=\psi_{e/o\,E}(x)$, and normalized as
\begin{equation}\label{eq:normalization}
\braket{\psi_{\alpha E}|\psi_{\beta E'}}=\delta\left(E-E'\right)\delta_{\alpha\beta},\quad \alpha,\beta\in\{e,o\}.
\end{equation}
The repulsive potentials have a dramatic effect on the continuum as they can keep the electron outside, or confine it within the barrier. As an example, Fig.~\ref{fig:2}.b shows the probability density of an odd scattering state at an energy close to that of the second odd bound state of a $2a$-wide box. Since in this case the potential barriers are only partially reflective, the bound state manifests itself as a resonance embedded in the continuum. The peak density within the region $[-a,a]$ is much larger than for $|x|>a$, which shows that the potential can transiently bind an electron. The interference mechanism that gives rise to this resonance confinement is the same as in the Fabry-P\'erot etalon of classical optics~\cite{Perot1899}.
It is useful to introduce also a second set of scattering states, $|\psi_{LE}^-\rangle$ and $|\psi_{RE}^-\rangle$, which satisfy incoming boundary conditions and represent a mono-energetic electron that, \emph{after} colliding with the potential (hence the minus), emerges on the left and on the right of the potential, respectively,
\begin{eqnarray}
	\psi_{LE}^-(x)&=&\frac{1}{\sqrt{\pi k}}\left\{
	\begin{array}{ll}
		e^{-ikx}+\beta e^{+ikx} \quad& x \le -a\\
		\gamma e^{-ikx} &  x \ge a
	\end{array}\right.\\
	\psi_{RE}^-(x)&=&\frac{1}{\sqrt{\pi k}}\left\{
	\begin{array}{ll}
		\gamma e^{ikx}& x \le -a\\
		\beta e^{-ikx}+e^{ikx}  \quad& x \ge a
	\end{array}\right.
\end{eqnarray}
where the complex numbers $\beta\equiv-\left(e^{-2i\eta_e}+e^{-2i\eta_o}\right)/2$ and $\gamma \equiv -\left(e^{-2i\eta_e}-e^{-2i\eta_o}\right)/2$ satisfy $|\beta|^2+|\gamma|^2=1$.
The left and right states, which turn into one another under the action of parity, $\Pi|\psi_{LE}^-\rangle=|\psi_{RE}^-\rangle$, can be expressed in terms of the even and odd scattering states, 
 \begin{equation}\label{eq:LRvseo}
       \psi_{L/R\,E}^-(x)=ie^{-i\eta_e}\psi_{eE}(x)\mp ie^{-i\eta_o}\psi_{oE}(x).
\end{equation}

\paragraph{Dipole transition matrix elements} 
To evaluate the interaction of a charged particle in the model potential with an external time-dependent electric field, we must compute the dipole matrix for both bound-free and free-free transitions. 
The only bound-free transition in the present 1D model is the one between the even ground state and the odd scattering states. In velocity gauge,
\begin{equation}
\mu_{E,g}=\braket{\psi_{oE}|p|\psi_g}=-i\int_{-\infty}^{\infty}\hspace{-5pt}dx \,\,\psi_{oE}(x)\frac{d}{dx}\psi_g(x).
\end{equation}
Under the assumption that the ground state wavefunction is negligible beyond $|x|>a$, the dipole integral can be evaluated using the analytical expression for both the initial and final wavefunctions in the inner region only,
\begin{eqnarray}
	\mu_{E,g}&=& -\frac{2iA_o(k)A_g\kappa k}{\sqrt{\pi k}(\kappa^2+k^2)}.
	 \label{eq:one-photon-mx}
\end{eqnarray}
Since neither the bra nor the ket are normalizable, the free-free dipole matrix element comprises both a delta and an \emph{off-shell} component,
\begin{eqnarray}
	&&\braket{\psi_{eE} |\frac{d}{dx}|\psi_{oE'}}=
	\frac{\mathcal{P}}{E'-E}\,\braket{\psi_{eE}|\left[d/dx,H_0\right]|\psi_{oE'}}\nonumber\\
	&&+\delta(k-k')\,\pi\,\sin\left\lbrack\eta_e(k)-\eta_o(k')\right\rbrack A_e(k)A_o(k'),\label{eq:ccdme}
\end{eqnarray}
where $\mathcal{P}$ indicates the principal part, and we have used the fact that the bra and ket are eigenfunctions of the field-free Hamiltonian to factor out the matrix element of $\left[d/dx,H_0\right]$, which is regular. The delta component is computed from the asymptotic expression of the scattering states. The regular factor in the \emph{off-shell} component can be readily shown to have the following expression,
\begin{equation}
\label{eq:td-cc}
\begin{split}
&\braket{\psi_{eE}|\left[d/dx,H_0\right]|\psi_{oE'}}=A_e(k)A_o(k') k k^\prime\times\\
    &\times \left\lbrace
    \cos\left[ka+\delta_e(k)\right]\cos(k' a)
    -\cos\left[\delta_e(k)\right]\right\rbrace-\\
    &-k k^\prime\cos\left[ka+\eta_e(k)\right]\cos\left[k^\prime a+\eta_o(k')\right].
\end{split}
\end{equation}	
The bound-free and free-free dipole matrices can be used to compute the one- and two-photon transition matrix elements. These transition matrices are used in turn to compute the finite-pulse multi-photon integrals to determine the one- and two-photon perturbative ionization amplitudes~\eqref{eq:one-photon} and~\eqref{eq:two-photon}.

\subsection{Numerical 1D model}
In the numerical version of the model, both the potential and the wavefunction are tabulated on a grid. Whereas the singular potential $V(x)$ of Eq.~\eqref{eq:pot_analy} cannot be represented in this context, it can be approximated by a potential $V_N(x)$ in which the delta distributions are replaced with narrow Lorentzian functions $L_\Delta(x)$,
\begin{equation}
	V_N(x)=V^N_+L_\Delta(x+a)+V^N_+L_\Delta(x-a)-V^N_-L_\Delta(x)\label{eq:pot_num},
\end{equation}
where $L_\Delta(x)=\Delta/[2\pi(x^2+\Delta^2/4)]$, with $\Delta\ll a$. 
The bound $\psi_g$ and scattering states $\psi_{e/o E}$ of the eigenvalue problem $H_0\psi = E\psi$ with the potential~\eqref{eq:pot_num} are computed  numerically using the renormalized Numerov method \cite{Johnson1977,Johnson1978}.
The scattering functions outside the potential barrier have the same form as the one given in~\eqref{eq:phi_even} and \eqref{eq:phi_odd} for $|x|>a$. 
The constant parameters $V^N_+$ and $V^N_-$, however, may not coincide with $V_+$ and $V_-$ if the energy difference between the bound and resonant states in the NM is to match that in the AM. The values of $\eta_{e/o}(k)$ are computed numerically by matching the eigenfunction of $H_0$ to the form \eqref{eq:phi_even} and \eqref{eq:phi_odd} at a point $x_M$ outside the potential barriers. 

\paragraph{Dipole transition matrix elements}
In the NM, the bound-free transition matrix elements are computed by numerical integration. For the free-free dipole matrix elements, the delta components \emph{on shell} is dictated exclusively by the asymptotic behavior of the continuum wave functions and hence it has the same form as the \emph{on shell} component of the free-free matrix element in the AM in~\eqref{eq:ccdme}.
The \emph{off-shell} component of the free-free dipole matrix element in the NM is computed by partitioning the integral $\int^{\infty}_{-\infty}\psi_{eE}(x)\psi'_{oE'}(x)dx$ into a short-range ($|x|<x_M$) and a long-range part, where $x_M$ is chosen so that the potential barrier is negligible for $|x|>x_M$. The short-range part, $2\int^{x_M}_{0}\psi_{eE}(x)\psi'_{oE'}(x)dx$, is evaluated numerically,
whereas the outer component can be expressed as the following boundary term
\begin{eqnarray}
&&\int^{\infty}_{x_M}\hspace{-3pt}\psi_{e E}\frac{d\psi_{oE'}}{dx}\,dx=\frac{k'}{2}\left(E-E^\prime\right)\times\nonumber\\ 
&\times\big\lbrace& k\cos\left[kx_M+\eta_e(k)\right]\cos\left[k^\prime x_M+\eta_o(k^\prime)\right]\nonumber\\
&+&k^\prime\sin\left[kx_M+\eta_e(k)\right]\sin\left[k^\prime x_M+\eta_o(k^\prime)\right]\big\rbrace.\label{eq:boundaryCorrection}
\end{eqnarray}

In the NM, the TDSE is solved in the length gauge and using finite differences with a three-point formula to evaluate the kinetic energy term. The TDSE solution is propagated forward in time using the Crank-Nicolson method~\cite{Ilchen2017}. The ionization amplitude is eventually extracted from the wavefunction by projecting it on the scattering states of the system, as $\mathcal{A}_{\alpha E\leftarrow g}=\langle \psi_{\alpha E}|\Psi(t)\rangle$.

\subsection{Physical observables}
The key observable in the 1-2 RABBITT scheme is the photoemission asymmetry, $\Sigma(E)$, defined here as the difference between the probability of photoemission to the left and to the right, per unit of energy,
\begin{equation}
\Sigma(E) = \mathcal{P}_L(E) - \mathcal{P}_R(E).
\label{eq:asymmetry}
\end{equation}
The directional photoelectron distributions are the square of the corresponding photoemission amplitudes, $\mathcal{P}_{L/R}(E)=|\mathcal{A}_{L/R\,E\gets g}|^2$, where $\mathcal{A}_{L/R\,E\gets g}=\braket{\psi^-_{L/R\,E}|\Psi_I(t)}$ are obtained by projecting $\Psi(t)$, after the end of the pulse. In turn, using~\eqref{eq:LRvseo}, the photoemission amplitudes to the left and to the right can be expressed in terms of the amplitudes to the even and odd scattering states, $\mathcal{A}_{oE\gets g}=\langle\psi_{oE}|\Psi_I(t)\rangle$ and $\mathcal{A}_{eE\gets g}=\langle\psi_{eE}|\Psi_I(t)\rangle$,
\begin{equation}
     \mathcal{A}_{L/R\, E\gets g}=\,-ie^{+i\eta_e}\,\mathcal{A}_{e E\gets g}\,\pm\, ie^{+i\eta_o}\,\mathcal{A}_{oE\gets g}.
\end{equation}
The even (odd) amplitudes result from the exchange, from the ground state, of an even (odd) number of photons. To lowest order, therefore, the left and right ionization amplitudes have the simplified form
\begin{equation}
     \mathcal{A}_{L/R\, E\gets g}\simeq\,-ie^{+i\eta_e}\,\mathcal{A}_{e E\gets g}^{(2)}\,\pm\, ie^{+i\eta_o}\,\mathcal{A}_{oE\gets g}^{(1)}.
\end{equation}
To the same order, the total ionization probability is
\begin{equation}
\mathcal{P}_{tot}(E)\simeq|\mathcal{A}^{(2)}_e(E)|^2+|\mathcal{A}^{(1)}_o(E)|^2,\label{eq:ionization_prob}
\end{equation}    
whereas the left-right asymmetry is
 \begin{equation}
\Sigma(E) \simeq -4\,\Re e\,[\mathcal{A}_o^{(1)*}(E)\,\mathcal{A}_e^{(2)}(E)\,e^{i(\eta_e-\eta_o)}],
 \label{eq:asym}
\end{equation}     
where $\Re e\, z$ is the real part of $z$. Equations~\eqref{eq:ionization_prob} and \eqref{eq:asym} show that the interference between one- and two-photon ionization pathways manifests itself in the photoemission asymmetry only.

To compute $\Sigma(E)$ perturbatively in the AM, it is necessary to evaluate the two-photon ionization amplitude \eqref{eq:two-photon-matrix}.
The two-photon matrix element in velocity gauge,
\begin{eqnarray}
\mathcal{M}_{eE\gets g}^{(2)}(\omega)&=&\mathcal{P}\int\frac{\langle \psi_{eE}|p|\psi_{oE'}\rangle\langle \psi_{oE'}|p|\psi_{g}\rangle}{E_g+\omega-E'}dE'-\nonumber\\ \label{eq:two-photon-matrix-2}
&-&i\pi{\langle \psi_{eE}|p|\psi_{o\,E_g+\omega}\rangle\langle \psi_{o\,E_g+\omega}|p|\psi_{g}\rangle},
\label{eq:two-photon-matrix-2}
\end{eqnarray}
has a principal-part $\mathcal{P}$ integral that we compute numerically from the dipole matrix elements, which are analytically known. The first term on the right hand side of \eqref{eq:two-photon-matrix-2}, which accounts for the contribution of the virtual excitations in the two-photon transition, is purely real, whereas the second term, which accounts for the sequential part of the excitation, is purely imaginary. Therefore, the phase $\varphi^{(2)}\approx\arg\left[\mathcal{M}_{eE\leftarrow g}^{(2)}(\omega_{\IR})\right]$ depends on the ratio between the sequential and virtual components of the transition. Notice how in the soft-photon approximation~\cite{Galan16}, which is sometimes used to model multiphoton processes, the ionization continua are not coupled \emph{off shell}, and hence only the principal part contributes. In general, however, the continuum-continuum coupling does have off-diagonal terms. In the present case, the most relevant \emph{off-shell} contribution is due to the intermediate resonance. Finally, the two-photon ionization amplitude is computed using \eqref{eq:two-photon} and the left-right asymmetry is subsequently obtained using \eqref{eq:asym}.

\section{Results and discussion} \label{sec:Results}
In this section we discuss the results of the 1-2 RABBITT pump-probe ionization scheme applied to the ground state of both the analytical~\eqref{eq:pot_analy} and the numerical~\eqref{eq:pot_num} model. In both cases the two repulsive barriers are set at $10$~a.u. from the origin ($a=10$~a.u.). The potential strengths are chosen as $V_+=2.5$~a.u. and $V_-=0.5$~a.u., in the analytical model [see \eqref{eq:pot_analy}], whereas in the numerical model they are set to $V^N_+=3.1105$~a.u. and $V^N_-=0.5$~a.u. [see \eqref{eq:pot_num}], with a width of the Lorentzian potentials of $\Delta=0.05$~a.u. With these choices, the two models have similar ground state energies, $E_g=-0.125$~a.u. (AM) and $E_g=-0.1112$~a.u. (NM), as well as comparable positions and widths of the resonant states.

\subsection{Scattering phase and dipole matrix}
As discussed in the previous section, the scattering states of the AM are analytically known. To illustrate their essential features and compare them with the properties in the NM, we plot in Fig.~\ref{fig:phase-shifts} the total resonant phaseshift for AM and NM, from which we have subtracted the background phase $\eta_{\mathrm{bg}}=-2ka$, to highlight the resonant features: $\eta_{\mathrm{tot}}=\eta_e+\eta_o-\eta_{\mathrm{bg}}$. The resonant phase shift encodes the information needed to predict how the motion of a wavepacket in the potential differs from a reference free wavepacket. The derivative of $\eta_{\mathrm{tot}}(E)$ with respect to the energy is proportional to the averaged group delay experienced by a wave packet as it is transmitted or reflected by the potential. The steep $\pi$ jumps, in particular, indicate a long confinement, and hence they reveal the presence of resonant states. Due to the difference in the definition of the potential, the phase shifts $\eta_{e/o}(k)$ in the NM are not expected to exactly coincide with those in the AM. Nevertheless, as the figure shows, they do compare well with each other. 
\begin{figure}[t]
\includegraphics[width=\columnwidth]{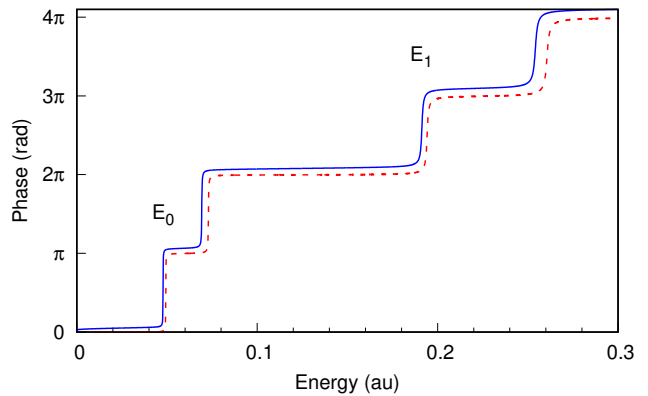}	
\caption{Scattering phase shift for AM (blue solid line) and NM (red dotted line). Also shown the position of the two lowest odd resonant state.}
\label{fig:phase-shifts}
\end{figure}

The prominent role of resonances in the ionization of the ground state, in the present model, and the good agreement between the AM and NM, is apparent in the dipole transition matrix element $\mu_{E,g}$ [compare with Eq.~\eqref{eq:one-photon-mx}], which is shown in Fig.~\ref{fig:bc-dipole-matrix}.
\begin{figure}[b]
\includegraphics[width=1.\columnwidth]{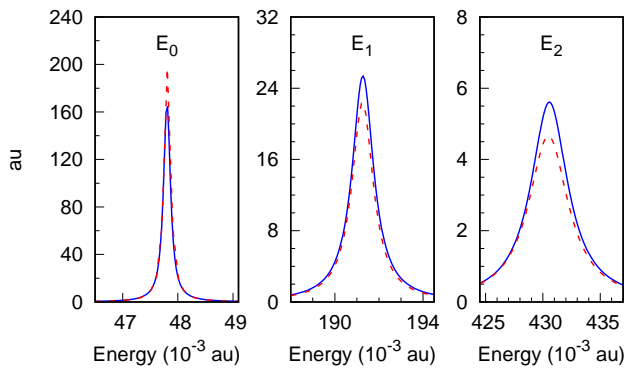}
\caption{\label{fig:bc-dipole-matrix} Square module of the bound-free dipole matrix element, as a function of the photoelectron energy calculated in AM (blue solid line) and NM (red dotted line).}
\end{figure}  
Since the position of the resonances in the two models do not exactly coincide, the NM calculations in the three panels of Fig.~\ref{fig:bc-dipole-matrix} have been shifted to lower energies by $0.0015$~a.u., $0.0029$~a.u., and $0.0039$~a.u., respectively, to better highlight the similarity of the resonant peaks' shape and amplitude in the two calculations. Notice that the scale differ in each panel: resonance widths increase with energy.

\begin{figure}[t]
\includegraphics[width=1.\columnwidth]{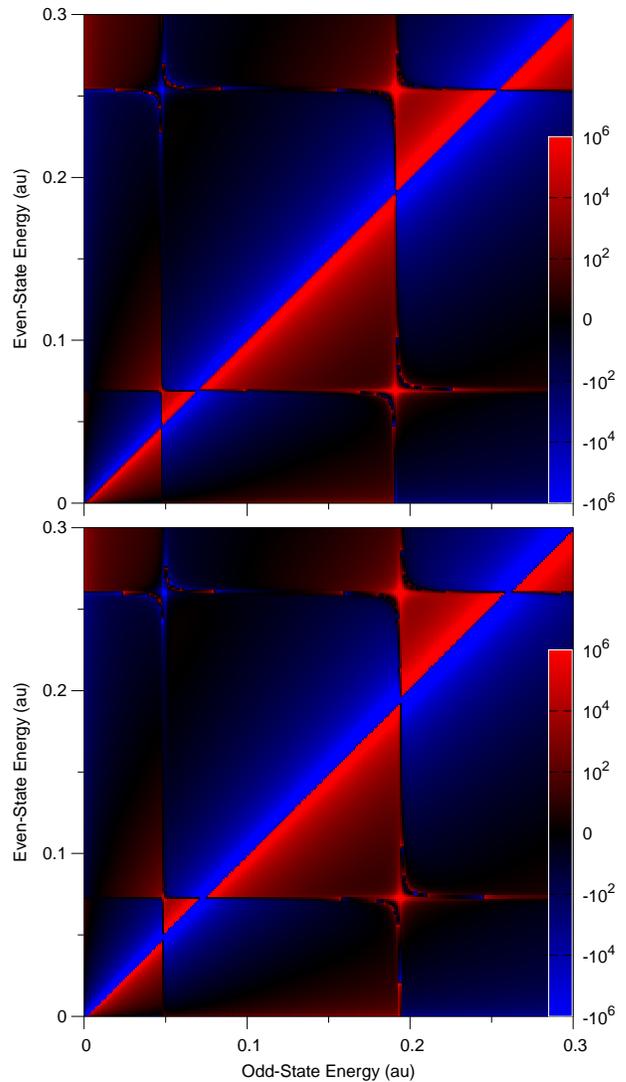}
\caption{\label{fig:cc-dipole-matrix} Free-free transition dipole moments as a function of the even-state and odd-state energies for (a) AM and (b) NM.}
\end{figure}
Regarding the free-free transition dipole moments~\eqref{eq:ccdme}, we have verified that those computed in the NM are independent of the specific choice of $x_M$ in~\eqref{eq:boundaryCorrection}, as long as $x_M$ lies outside the effective range of the numerical potential barrier. The free-free dipole matrix elements for the two models are shown in Fig.~\ref{fig:cc-dipole-matrix} and found to be in excellent agreement with each other. In the figure, we observe the clear signature of the odd and even resonances, as well as the characteristic $(E-E')^{-1}$ singularity.

\subsection{Ionization amplitudes}
Let us consider a pump-probe scheme with electric field
\begin{equation}
E(t)=E_{\XUV}(t)+E_{\IR}(t,\tau),
\end{equation}
where the pump, which comprises two consecutive XUV harmonics, overlaps with a delayed IR probe pulse, 
\begin{eqnarray}
E_{\XUV}(t)&=&E_{\XUV}F_{\XUV}(t)\left[\sin(\omega_{2n}t)+\sin(\omega_{2n+1}t)\right],\nonumber\\
E_{\IR}(t,\tau)&=&E_{\IR}F_{\IR}(t-\tau)\sin[\omega_{\IR}(t-\tau)].
\end{eqnarray}
The two harmonics have equal maximum field amplitude $E_{\XUV}=0.004$~au, a Gaussian envelope $F_{\XUV}(t)$ with ${\rm FWHM}\approx 12$~fs, and $\Delta \omega=\omega_{2n+1}-\omega_{2n}=\omega_{\IR}$.
The second harmonic $\omega_{2n+1}$ is tuned to the second odd resonance $E_1$ (see Fig.~\ref{fig:phase-shifts}), in both the AM and NM calculations. 
The probe IR has a frequency $\omega_{\IR}=0.051$~au, with ${\rm FWHM}\approx 57$~fs, and is delayed 
with respect to the maximum of the harmonics' envelope by a time-delay $\tau$. The XUV and IR pulses are shown in Fig.~\ref{fig:pulse}. 
 \begin{figure}[t]
\includegraphics[width=1.\columnwidth]{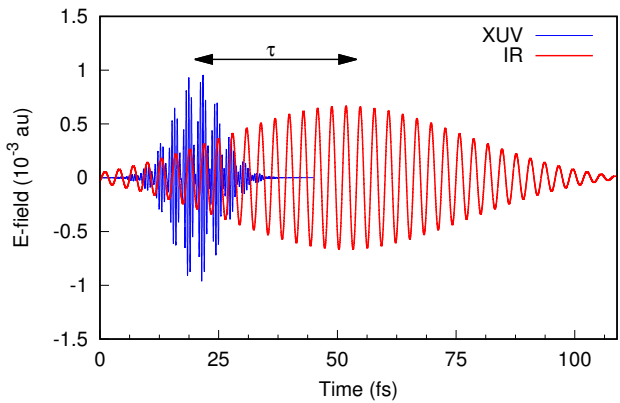}
\caption{\label{fig:pulse} XUV and IR pulses used in the study. Positive time delay $\tau$ indicates that the XUV harmonics arrive before the peak of the IR envelope.}
\end{figure}
\begin{figure}[b]
\includegraphics[width=1.\columnwidth]{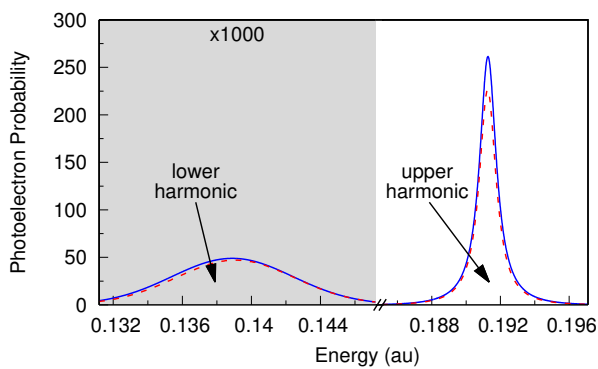}
\caption{\label{fig:tot-specs} Photoelectron distribution when using the XUV pump pulses only: AM (solid line), NM (dashed line). The ionization probability near the lower harmonic (grey region) has been multiplied by a factor of $1000$ to make it visible on the scale of the upper harmonic.}
\end{figure}

According to Eq.~\ref{eq:one-photon}, the one-photon ionization amplitude from the ground state, due to the XUV harmonics, is
 $ \mathcal{A}_{oE_f\leftarrow g}^{(1)} = -i\tilde{A}(\omega_{fg})\,\mu_{E_f,g}$, where $E_f$ is the final electron energy.
The ionization probability, $\mathcal{P}(E_f)=|\mathcal{A}_{oE_f\leftarrow g}^{(1)}|^2$, computed for the AM and the NM in the absence of the IR field are in excellent agreement, as shown in Fig.~\ref{fig:tot-specs}.
The ionization probability at the upper harmonic (resonance), induced by the higher harmonic $\omega_{2n+1}$, is orders of magnitude larger than that at the lower off-resonance harmonic $\omega_{2n}$, indicating that the one-photon background photoemission is negligible. This circumstance is to be expected since the ground state wavefunction, which is confined inside the potential barrier, has nearly zero overlap with the off-resonance free functions, which barely penetrate inside the barrier. This is also the reason why the resonances in our calculations have the symmetric Lorentzian shape characteristic of systems with negligible direct-ionization cross sections. 
If the product $a\,V_-$ were much smaller, the ground state would eventually extend outside the confinement barrier, and the resonances  exhibit an asymmetric lineshape analogous to the celebrated Fano profile~\cite{Fano1961}.
\begin{figure}[b]
\includegraphics[width=1.\columnwidth]{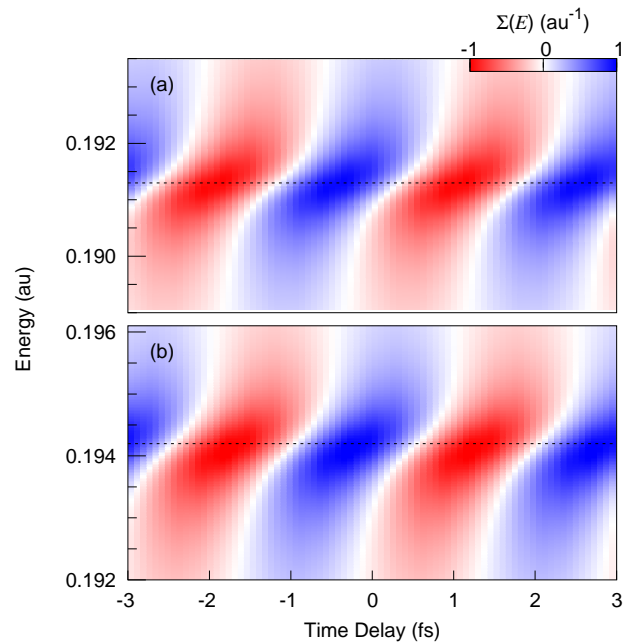}
\caption{\label{fig:8-res} Left-right asymmetry $\Sigma(E)$ as a function of the time-delay $\tau$ and the photoelectron energy, computed for AM (PT) (a) and NM (TDSE) (b), near the resonance $E_1$ where the dashed line indicates the center of the upper harmonic for each model.}
\end{figure}

\subsection{Two-Photon Ionization amplitude and asymmetry}
The left-right asymmetry \eqref{eq:asym} depends on the phases of the one- and two-photon ionization amplitudes 
\begin{eqnarray}
\arg[\mathcal{A}_o^{(1)}(E)]&=&\varphi_{\XUV},\\
\arg[\mathcal{A}_e^{(2)}(E)]&=&\varphi_{\XUV}+\omega_{\IR}\tau+\varphi^{(2)},
\end{eqnarray}
where $\varphi_{\XUV}$ is the XUV phase, which is constant across the two harmonics, and $\varphi^{(2)}$ is the two-photon phase. The particular value of the relative phase between the two harmonics, which may depend on the process used to generate them, is irrelevant. For an IR pulse covering many cycles, 
as it is the case in our study, $\varphi^{(2)}\approx\arg\left[\mathcal{M}_{eE\leftarrow g}^{(2)}(\omega_{\IR})\right]$.
Therefore, the asymmetry~\eqref{eq:asym} beats at the same frequency as the IR field, as a function of the time delay $\tau$,
\begin{equation}
\Sigma(E) =\Sigma_0(E)\cos(-\omega_{\IR}\tau+\varphi^{(2)}+\eta_e-\eta_o),
\label{eq:asym-dev}
\end{equation}
where $\Sigma_0(E) = -4|\mathcal{A}^{(1)}_{oE\gets g}\mathcal{A}_{eE\gets g}^{(2)}| $.
\begin{figure}[t]
\includegraphics[width=1.\columnwidth]{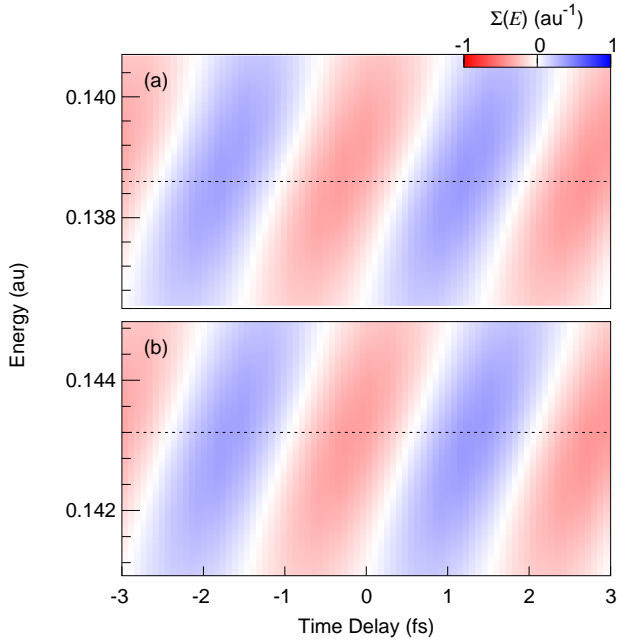}
\caption{\label{fig:8-sid} Same as in Fig.~\ref{fig:8-res} but at the lower harmonic and the dashed line now indicates the center of the lower harmonic for each model.}
\end{figure}
The NM photoionization asymmetry is readily obtained by projecting the final wavefunction $\Psi(t)$ from the TDSE simulation onto the left $\psi_{LE}$ and right $\psi_{RE}$ scattering states and using \eqref{eq:asymmetry}. 
The left-right asymmetry, recorded as a function of the time delay between the XUV-pump and IR-probe, is plotted in Figs. \ref{fig:8-res} and \ref{fig:8-sid}, for both AM (PT) and NM (TDSE).
 The figures demonstrate the predicted $\omega_{\IR}$ beating for each harmonic. In Fig.~\ref{fig:8-res} we present the asymmetry beating at the upper harmonic, near the resonance, whose phase experiences a dramatic excursion across the resonance. This result is easily understood from \eqref{eq:asym-dev}: whereas the phase $\varphi^{(2)}+\eta_e$ varies smoothly across the resonance, $\eta_o$ undergoes the same rapid excursion visible already in Fig.~\ref{fig:phase-shifts}. From Fig.~\ref{fig:8-res}, therefore, it is possible to extract the pure one-photon scattering phase associated with that resonant state.
 In contrast to the asymmetry beating at the resonant harmonic, the phaseshift at the lower harmonic in Fig.~\ref{fig:8-sid} shows a much less pronounced modulation, similar to the one that is accessible with traditional RABBITT measurements~\cite{Galan16,Kotur2016,Gruson2016}.
 The main phase variation is now due to $\varphi^{(2)}$, which encodes the resonance phase through the first XUV photon absorption, distorted by the free-free transition from the upper to the lower harmonic.
 Even in this case the PT and TDSE are in excellent agreement.
\begin{figure}
\includegraphics[width=0.8\columnwidth]{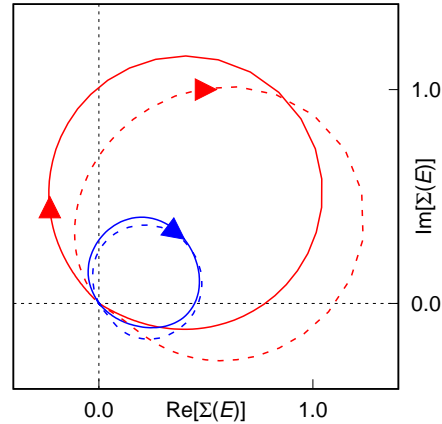}
\caption{\label{fig:9} Averaged Fourier transform of the left-right asymmetry, with respect to the time delay, $\bar{\Sigma}(E)$ in the complex plane at the resonance (large circles, red) and at the lower harmonic (small circles, blue), for AM (solid lines) and NM (dashed lines). The arrows indicate the evolution of the trajectory with increasing energy.}
\end{figure}
To extract the phaseshift from the asymmetry beating at the upper harmonic we computed the Fourier transform
\begin{equation}
 \tilde{\Sigma}(E;\omega)=\frac{1}{\sqrt{2\pi}}\int^{\infty}_{-\infty}\xi(\tau)\,\,\Sigma(E;\tau)\,e^{-i\omega\tau}\,d\tau,
\end{equation}
where $\xi(\tau)$ is a smooth window function that allows us to limit the overlap between the FT of the background, centered at $\omega = 0$, with that of the beating signal, which is centered at $\omega=\omega_{\IR}$,
\begin{eqnarray}
\xi(\tau)=\frac{1}{2}\left[1-{\rm erf}(2\tau+T_\xi)\,{\rm erf}(2\tau-T_\xi)\right],
\end{eqnarray}
${\rm erf(x)}=2/\sqrt{\pi}\int_0^x\exp(-t^2)dt$, and $T_{\xi}=9T_{\IR}$, with $T_{\IR}=2\pi/\omega_{\IR}$ being the IR period.
Finally, $\tilde{\Sigma}$ is integrated across a spectral width $\Delta\omega\approx0.02$~a.u. around $\omega_{\IR}$,
\begin{eqnarray}
\bar{\Sigma}(E)=\int_{\omega_{\IR}-\Delta\omega/2}^{\omega_{\IR}+\Delta\omega/2}\tilde{\Sigma}(E;\omega)\,d\omega.
\end{eqnarray}
The resulting complex amplitude, $\bar{\Sigma}(E)$, are plotted for both models in Fig.~\ref{fig:9}.
The complex amplitudes exhibit the characteristic jump of $\pi$ as the energy crosses the resonance while at the sideband the jump of $\pi$ is modulated by the emission of an IR-photon from a pulse of finite length. 
\begin{figure}
\includegraphics[width=\columnwidth]{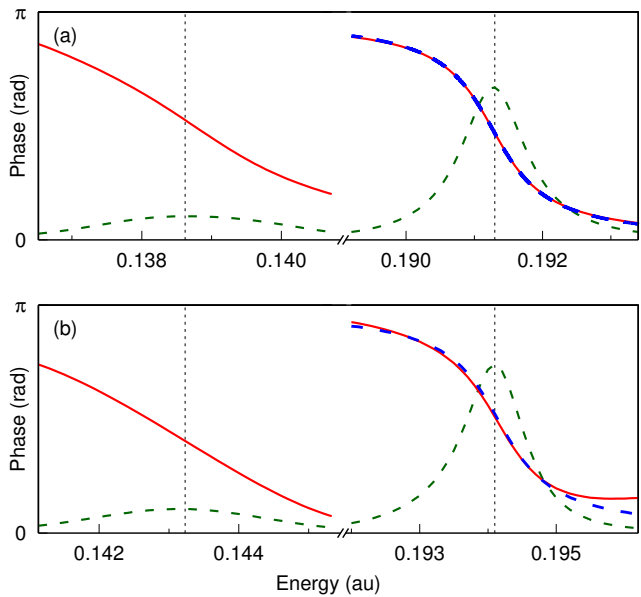}
\caption{\label{fig:ion-asymm} Retrieved scattering phase $\eta_o^{ret} $obtained from the computed phase of $\bar{\Sigma}(E)$ (red solid line) and scattering phaseshift  $\eta_o$ (blue dashed line) near the resonance plotted for AM (PT) in (a) and NM (TDSE) in (b).  The asymmetry amplitude $|\bar{\Sigma}(E)|$ obtained from both models is also shown (green dashed line).}
\end{figure}
The resulting phases and amplitudes are plotted separately in Fig.~\ref{fig:ion-asymm}. We also compare the scattering phase at resonance retrieved from $\bar{\Sigma}(E)$ with the scattering phases $\eta_o$. To highlight the close similarity between the scattering phase and the phase retrieved from the pump-probe spectrum we have shifted the latter by a constant $\eta_c$. The striking agreement between the two phases demonstrates that the use of consecutive harmonics of the probe pulse holds the key to direct measurement of Wigner delays~\cite{Wigner1955,Pazourek2015}.
To extract the resonance position and width, the phase shift $\eta_{\mathrm{tot}}(E)$ is fitted around the resonance to the argument of the Breit-Wigner amplitude and a quadratic background $\delta_{\mathrm{bg}}(E)$ of the form: 
\begin{equation}
\eta_{\mathrm{tot}}(E)=\acot[2(E-E_r)/\Gamma]+ \delta_{\mathrm{bg}}(E).
\end{equation}
The resonance width, $\Gamma$, and position, $E_r$, obtained from the scattering phaseshift are listed in Tab. \ref{tab:1} along side $\Gamma^{ret}$ and $E^{ret}_r$ fitted from $\eta^{ret}_o$.
\begin{table}[hbtp!]
\caption{\label{tab:1} Comparison of resonance position and width obtained directly from $\eta_o$ or from the retrieved  phaseshift  $\eta^{ret}_o$. All values are given in atomic units.}
\begin{ruledtabular}
\begin{tabular}{ccccc}
&$E_r$&$\Gamma\times10^3$&$E^{ret}_r$&$\Gamma^{\rm ret}\times10^3$\\
\hline
AM & 0.1913 & 1.184 & 0.1913 & 1.181 \\
NM & 0.1942 & 1.208 & 0.1942 & 1.183 \\
\end{tabular}
\end{ruledtabular}
\end{table}

\section{Conclusion} \label{sec:CONCLUSION}
We have presented 1-2 RABBITT, a new perturbative attosecond interferometric spectroscopy  in which two consecutive isolated harmonics of the probe pulse are used, instead of full comb of odd harmonics, as it is the case in traditional RABBITT. Using a 1D potential that supports bound as well as autoionizing states, we have shown how the new technique allows us to retrieve full phase information of the photoelectron-emission amplitude without the measurement-induced distortion inherent to traditional RABBITT. This feature results from the direct interference between one- and two-photon transition pathways. Furthermore, 1-2 RABBITT is also able to quantify the measurement-induced distortion due to the exchange of a probe photon. These findings have general validity and can be applied to study autoionizing states in real systems, such as rare-gas atoms. Specific generation schemes in which the APT comprises isolated consecutive harmonics of the probe, such as when the probe is the second harmonic of the fundamental frequency, or when the APT is generated with bicircular light pulses, as well as their application to real atomic systems, is beyond the scope of the present work and will be the subject of future investigations.

\section*{ACKNWLEDGMENTS}
This work was supported by the United States National Science Foundation under NSF Grant N$^\circ$~PHY-1607588, and by the UCF in-house OR grant Acc. N$^\circ$~24089045.

\bibliography{biblio.bib}

\end{document}